# Highly efficient spin-to-charge current conversion at room temperature in strained HgTe surface states

P. Noel[1], C. Thomas[2,] Y. Fu[1], L. Vila[1], B. Haas[3], P-H. Jouneau[3], S. Gambarelli[4], T. Meunier[5], P. Ballet[2], J.P. Attané[1]

1 Univ. Grenoble Alpes, CEA, CNRS, Grenoble INP, INAC, SPINTEC, F-38000 Grenoble, France

2 Univ. Grenoble Alpes, CEA, LETI, MINATEC campus, F38054 Grenoble, France

3 CEA, INAC-MEM, 38054 Grenoble, France

4 CEA, Institut Nanosciences et Cryogénie, SyMMES F-38000 Grenoble, France

5 CNRS, Institut NEEL, 38042 Grenoble, France

We report the observation of spin-to-charge current conversion in strained mercury telluride at room temperature, using spin pumping experiments. The conversion rates are found to be very high, with inverse Edelstein lengths up to 2.0 ± 0.5 nm. The influence of the HgTe layer thickness on the conversion efficiency has been studied, as well as the role of a HgCdTe barrier inserted in-between the HgTe and NiFe layers. These measurements, associated to the temperature dependence of the resistivity, allows to ascribe these high conversion rates to the spin momentum locking property of HgTe surface states.

Conventional spintronics is based upon the use of magnetic materials to manipulate spin currents [1]. Such a manipulation can be achieved by harnessing the spin-orbit coupling in non-magnetic materials. For instance, the Spin Hall Effect [2] permits to convert charge currents into spin currents in the bulk of heavy metals. It has also been recently demonstrated that higher conversion rates can be obtained by using two-dimensional electron gas with high spin-orbit coupling, in Rashba interfaces [3] or at topological insulator (TI) surfaces [4,5,6,7]. As a consequence, the use of Rashba interfaces, such as Ag/Bi [8,9] or STO/LAO [10], and of various TIs [11,12,13,14,15], is generating a growing attention in spintronics.

The main interest of TIs lies in their surface states, which possess a linear Dirac-like energy dispersion, and in the perpendicular locking between spin and momentum [4,5,6,7]. A flow of electric current in the two-dimensional electron gas gives rise to a perpendicular spin accumulation, this effect being known as the Edelstein Effect [16], while the reverse spin-to-charge conversion phenomenon is known as the Inverse Edelstein Effect (IEE) [17].

The conversion has been observed in various Bi-based TIs such as $Bi_2Se_3$ [11, 12], BiSbTeSe [14] or Sn-doped BiTeSe [15]. Although large spin torque efficiencies have been measured in $Bi_2Se_3$ [18,19], this system exhibits at room temperature relatively low surface related spin-to-charge conversion rates [20]. Also, due to intrinsic doping by selenium vacancies, $Bi_2Se_3$ presents bulk metallic states. In bulk-insulating BiSbTeSe and Sn-doped BiTeSe, a high conversion rate has been observed by spin pumping, but only at low temperature. The conversion has also been observed in strained α-Sn [13], but in order to preserve the Dirac cone a highly conductive layer has to be inserted between the ferromagnetic layer and the surface states, inducing a large magnetic damping which could limit its interest for spin-orbit torques.

In that context, strained HgTe is a TI expected to exhibit high conversion rates, as it is characterized by a very high mobility of its surface states [21]. Moreover, HgTe/CdTe is an archetypal topological insulator [22], compatible with electronic [23,24] and optoelectronic applications [25]. Beyond classical spintronic applications such as current-induced magnetization switching, the conversion of charge currents into spin currents in HgTe could thus pave the way for the use of spins as data carriers in all-semiconductor-based technologies.

In this letter, we demonstrate the spin-to-charge current conversion at room temperature in strained HgTe. Using spin pumping, we measure a very high conversion efficiency. We show that the dependence of the conversion rate with the thicknesses of the HgTe and HgCdTe layers, associated to the temperature dependence of the resistivity, allows to ascribe this high conversion rate to the spin momentum locking at the surface states of HgTe.

Strained HgTe is known to be a TI with Dirac surface states [21,22 ,26]. The light hole band $\Gamma_{8,LH}$ is lying 0.3 eV above the $\Gamma_6$ band. Such an inverted band structure at the $\Gamma$ point results in topological surface states, robust to the presence of the heavy hole band $\Gamma_{8,HH}$ (cf. fig. 1a). Band gap opening and TI properties can then be induced in HgTe by applying a tensile strain, which lifts the degeneracy at the $\Gamma$ point. Experimentally, the tensile strain state can be achieved by growing HgTe on a substrate having a larger lattice constant, such as CdTe. In these conditions, the existence of a Dirac Cone at the free surface of HgTe has been confirmed by ARPES measurements [27].

Here, HgTe thin films have been grown by molecular beam epitaxy (the growth conditions are detailed in refs. [28,29]). After the deposition on a CdTe (001) substrate of a 200 nm thick CdTe buffer layer, a strained HgTe layer (10 nm to 80 nm thick) has been grown, immediately capped with a 5nm thick $Hg_{0.3}Cd_{0.7}Te$ insulating barrier to avoid any Hg desorption. After deposition, the thicknesses of both the HgTe and HgCdTe layers have been controlled by X Ray Reflectivity (XRR) on a PANalytical Empyrean X-Rays diffractometer, using the cobalt Kα1 wavelength of 1.789 Å (cf. figure 1b), the data being fitted using the Gen-X software [30]. As the XRR is performed ex-situ, the HgCdTe insulating barrier is partially oxidized. The presence of this CdTe-based oxide has been taken into account to fit the XRR data. The roughness estimated by XRR for the HgTe layer and HgCdTe capping was inferior to 0.5 nm. The crystallographic quality of the heterostructure and the sharpness of the HgTe/HgCdTe interface have also been controlled by High-Angle Annular Dark-Field (HAADF) imaging in a scanning transmission electron microscope (cf. Figure 1c) [31]. The associated intensity profile allows for the marking of the interface chemical boundaries between HgTe and $Hg_{0.7}Cd_{0.3}Te$. The interface width of 1.4nm has to be considered as an upper bound as the intensity profile is averaged over the 50-100nm thickness of the focused-ion-beam-prepared lamellae.

To perform spin pumping experiments, a 20 nm thick NiFe layer has been deposited ex-situ by evaporation. An Ar etching has been performed prior to the NiFe deposition, in order to remove the oxide layer, and eventually to modulate the thickness of the HgCdTe barrier.

After the deposition of the NiFe layer, the thicknesses of the NiFe and HgCdTe films have been controlled by XRR. The samples have then been cut into 0.4 mm wide and 2.4 mm long stripes, before being measured by conventional spin pumping ferromagnetic resonance experiments in cavity [32]. A static magnetic field $H$ has been applied in the plane of the sample, while a radio frequency field $h_{rf}$ at 9.68 GHz has been applied perpendicularly, thus leading the magnetization of the NiFe thin film to precess (cf. fig. 2a). The saturation magnetization $\mu_0 M_s$ of the NiFe layer extracted from the resonance field is found to be 1.02± 0.03 T for all the studied samples.

At the ferromagnetic resonance, a pure spin current flows from the NiFe/HgCdTe interface towards the HgTe layer [33]. This flow is evidenced by the increase of the Gilbert damping $\alpha$, revealing the extra magnetic relaxation channel that appears when adding the HgCdTe/HgTe/CdTe stack to the NiFe layer. The damping parameter α can be extracted from the linear dependence of the peak-to-peak linewidth $\Delta H_{pp}$ with the frequency $f$ of the *rf* field using stripline broadband FMR (cf. fig. 2b):

$$\Delta H_{pp} = \Delta H_0 + \frac{2}{\sqrt{3}}\left(\frac{2\pi}{\gamma}\right)\alpha$$

where $\gamma$ is the gyromagnetic ratio and $\Delta H_0$ the inhomogeneous contribution to the linewidth.

The extra Gilbert damping Δα due to the spin pumping is calculated by comparing the damping parameter of a reference NiFe(20 nm)//Si sample (α=6.33 ± 0.03×10⁻³) to those of the NiFe/HgCdTe/HgTe//CdTe samples. This extra damping is in the range of 0.1×10⁻³ to 2×10⁻³ for all studied samples, up to 10 times smaller than the damping induced by Pt or α-Sn [13]. Such low values underline the potential of HgTe for spin torque experiments, as the switching current is dominated by the $\alpha M_s^2$ term [34].

The extraction of the conversion efficiency has been done using the model proposed by Mosendz et al. [35]. The extra damping Δα is related to the effective spin mixing conductivity $G_{eff}^{\uparrow\downarrow}$, which expresses the global spin transmission:

$$G_{eff}^{\uparrow\downarrow} = \frac{4\pi M_s t_{NiFe}}{g\mu_B}\Delta\alpha$$

where $t_{NiFe}$ is the thickness of the NiFe layer (20 nm), $g$ is the Landé factor of NiFe, and $\mu_B$ is the Bohr Magneton.

At the ferromagnetic resonance, a spin current is appearing, directed vertically from the NiFe layer towards the strained HgTe. Its density, $J_s^{3D}$, can be written [33]:

$$J_s^{3D} = \frac{G_{eff}^{\uparrow\downarrow}\gamma^2\hbar {h_{rf}}^2}{8\pi\alpha^2}\left[\frac{4\pi M_s\gamma + \sqrt{(4\pi M_s\gamma)^2 + 4\omega^2}}{(4\pi M_s\gamma)^2 + 4\omega^2}\right]\frac{2e}{\hbar}$$

where $\omega$ is the angular frequency of the *rf* field. This pure spin current can then be converted into a transverse DC charge current I$_C$, by Inverse Spin Hall Effect (ISHE) or IEE. In the open circuit conditions presented in figure 2a, this charge current results in the appearance of a measurable voltage [32,33,35].

Figure 2c presents the ferromagnetic resonance signal, together with the spin pumping signal, for a 18.5 nm thick HgTe sample covered by a 1.6 nm thick HgCdTe layer. The measured voltage $V$ corresponds to the sum of two components, $V_{sym}$ and $V_{asym}$, symmetric and antisymmetric with respect to the resonance field. They can be extracted using a Lorentzian fit of the data. In the geometry presented in figure 2a, the contribution of the spin rectification effects to the signal can be easily eliminated by using the symmetries of the signal. Indeed, the IEE and ISHE contributions are symmetric with respect to the resonance field, contributing only to $V_{sym}$, whereas the anisotropic magnetoresistance contribution is antisymmetric. Moreover, the anomalous Hall effect and spin-Seebeck effect are symmetric with respect to the applied field (*i.e.*, $V(H) = V(-H)$), contrarily to the ISHE and IEE contributions [36].

As can be seen in figs. 2b and 2c, the signal is mostly symmetrical with respect to the resonance field, and its sign is well reversed when turning the sample by 180°, which implies that V(H) is dominated by the IEE or ISHE contributions, and that the anomalous Hall and spin-Seebeck effects are negligible. Note that the generated current is proportional to the precession angle, *i.e.*, to $h_{rf}^2$. In our experiments, $\mu_0 h_{rf} \sim 0.1\ mT$. As $h_{rf}$ can slightly vary from experiment to experiment (from 0.07 to 0.11 mT), a renormalization has been performed, for the sake of comparison between the samples. All the measured voltages and currents are given here for a reference excitation amplitude of $\mu_0 h_{rf} = 0.1 mT$.

The most striking result is the presence of a very efficient conversion at room temperature: the produced charge current *Ic* (up to 1.7 µA) is found to be much larger than what can be obtained with heavy metals (*e.g.*, 0.5 µA in a thick Pt sample [37,38]), and of the same order of magnitude as the highest value reported to our knowledge in a TI (2 µA in alpha-Sn [13], under the same $h_{rf}$ field).

Let us focus on the dependence of the charge current with the HgCdTe barrier thickness, shown in figs. 3a and 3b. Thin tunnel barriers, from 0.6 nm to 3 nm, allow obtaining higher currents than the direct contact between NiFe and HgTe. This enhancement may be due to the protection from proximity effects offered to the electronic states at the HgTe surface, and to the increase of the carrier lifetime [13]. Nonetheless, as the barrier thickness is increased the signal decreases, and vanishes for thick barriers [39]. This extinction confirms that the observed conversion does not occur at the NiFe/HgCdTe interface. The decrease of the signal with the barrier points toward a decrease of the electronic coupling through the insulating HgCdTe. Therefore, tuning appropriately the cap HgCdTe thickness enables the spin-to-charge conversion to be maximized.

To demonstrate that the conversion is due to the HgTe topological surface states, we studied the temperature dependence of the sheet resistance for different HgTe thicknesses (see fig. 3c). A resistance maximum is observed at around 50 K for the 18.5 nm thick HgTe layer. Its existence suggests the presence of two parallel channels of conduction, the first one corresponding to the bulk of HgTe, with a resistivity decrease when increasing the temperature, the second one corresponding to the metallic surface states, dominating the conductivity at low temperature.

When increasing the HgTe thickness up to 56 nm, the bulk contribution dominates the signal: the resistance keeps increasing at low temperature, without any signature of a metallic behavior. For a thinner (8.5 nm) sample, there is no overall decrease of the sheet resistance with increasing temperature, which indicates that for this sample the conductivity is dominated by the surface states.

Let us now estimate the spin-to-charge conversion factor, *i.e.*, the ratio of the obtained to the injected current densities. This value, which denotes the efficiency of the conversion from a spin current $J_s^{3D}$ (in A/m²) into a surface charge current $J_c^{2D}$ (in A/m), is known as the Inverse Edelstein length $\lambda_{IEE}$:

$$\lambda_{IEE} = \frac{J_c^{2D}}{J_s^{3D}} = \frac{I_C}{w J_s^{3D}}$$

where w is the sample width.

Fig. 3d shows the dependence of $\lambda_{IEE}$ with the HgTe thickness. If the dependence was due to the SHE in HgTe, we would expect a hyperbolic tangent increase with the HgTe thickness, *i.e.*, $\frac{I_c}{J_s^{3D}} \propto \tanh\frac{t}{2l_{sf}}$, t being the HgTe thickness and $l_{sf}$ its spin diffusion length [33]. The observed dependence is very different, with a large increase of $\lambda_{IEE}$ from t=8.5 nm to 26 nm, where the highest Inverse Edelstein length is obtained, and after which the efficiency drops.

In an ideal topological insulator, the inverse Edelstein length is equal to the product of the mean free path of the surface states $\lambda$ by the spin polarization of the surface states P [40]. Yet if the bulk is not perfectly insulating it acts as a shunt [14,15], and the inverse Edelstein length scales as:

$$\lambda_{IEE} = P\lambda \frac{R_b}{R_b + R_s}$$

where Rb and Rs are the sheet resistance for the bulk and surface states, respectively. This shunting explains the decrease of $\lambda_{IEE}$ for thick HgTe layers observed in Fig. 3d.

$\lambda_{IEE}$ also decreases at small HgTe thicknesses (*i.e.*, below 26 nm). We attribute this effect to the hybridization of the surface state of the upper and lower HgTe surfaces. Due to the overlap of the two surface wave functions, an electronic transport through states delocalized between the surfaces can be observed [41], where the spin degeneracy is restored. As a consequence, the spin-momentum locking properties, and thus the polarization and the spin-to-charge conversion efficiency are expected to progressively disappear as the HgTe thickness shrinks [42]. The thickness of 26 nm, below which the decrease related to the gap opening is observed, is comparable to what can be expected from theoretical calculations [31]).

An interesting feature of the spin pumping method is its ability to determine the chirality of the direction of the spin rotation around the Fermi circle. $\lambda_{IEE}$ is found in our experiment to be positive. According to Hall and ARPES measurements [27], the Fermi level is expected to be above the Dirac point. This indicates that the helical fermi contour is counter-clockwise in the upper part of the cone (as illustrated in fig. 1a), in accordance with theoretical predictions [43].

Beyond its sign, the amplitude of the conversion rate is noteworthy. The produced charge currents are large, in the µA, and $\lambda_{IEE}$ can reach a value of 2.0 ± 0.5 nm, comparable to that of alpha-Sn ($\lambda_{IEE}$=2.1 nm in ref.[13]), *i.e.*, the highest value recorded up to now at room temperature. Note that $\lambda_{IEE}$ can also be compared to the product of the SHE angle by the spin diffusion length $\theta_{SHE} l_{sf}$ [9]. In the case of conventional spintronic materials such as Pt, the reported value is 0.57 nm [44].

Such a high value can be explained by considering the high mobility µ of the surface states, which can be larger than $10^5$ cm²/V/s at 1.9K [21]. The mean free path of the surface states is related to the mobility [45] as

$$\lambda = \frac{\mu \hbar k_F}{e}$$

where $k_F$ is the wavevector at the Fermi level, µ the mobility and e the electron charge. Using the value $k_F$=3.10$^{-3}$nm$^{-1}$ obtained by ARPES measurements in similar samples [27], and a mobility of $10^5$cm²/V/s, this leads to an expected mean free path of 1.8 µm at low temperature. This value is one or two orders of magnitude larger than the mean free path of Bi-based TIs such as Bi$_2$Te$_3$ [45], BiSbTeSe [46] or Sn-doped BiTeSe [15] at low temperature. Even though the mean free path of HgTe is expected to decrease at room temperature, it might remain larger than those of Bi-based topological insulators, and thus lead to higher conversion rates.

To conclude, we observed at room temperature the spin-to-charge current conversion in the topological surface states of strained HgTe, with a counter-clockwise direction of the spin rotation, and very high conversion rates. The conversion can be optimized using a HgCdTe barrier. To obtain the highest conversion rate, the HgTe layer thickness has to be thick enough to decouple the top and bottom surface, but thin enough to avoid the electrical shunt by the bulk.

The gate dependence of the effect remains to be studied, and one can expect to enhance, or at least modulate, the conversion rate [10]. This degree of freedom, and the compatibility of the CdTe/HgTe system with standard processes in microelectronics, make it a very good candidate for applications based on spin-orbit torques, using either ferromagnetic metals or ferromagnetic semiconductors (*e.g.*, Cr-doped CdTe [47]). Beyond the topic of magnetization switching by spin-orbit torques, the ability to manipulate spins in HgTe also represents a step towards the development of an all-semiconductor spintronics.

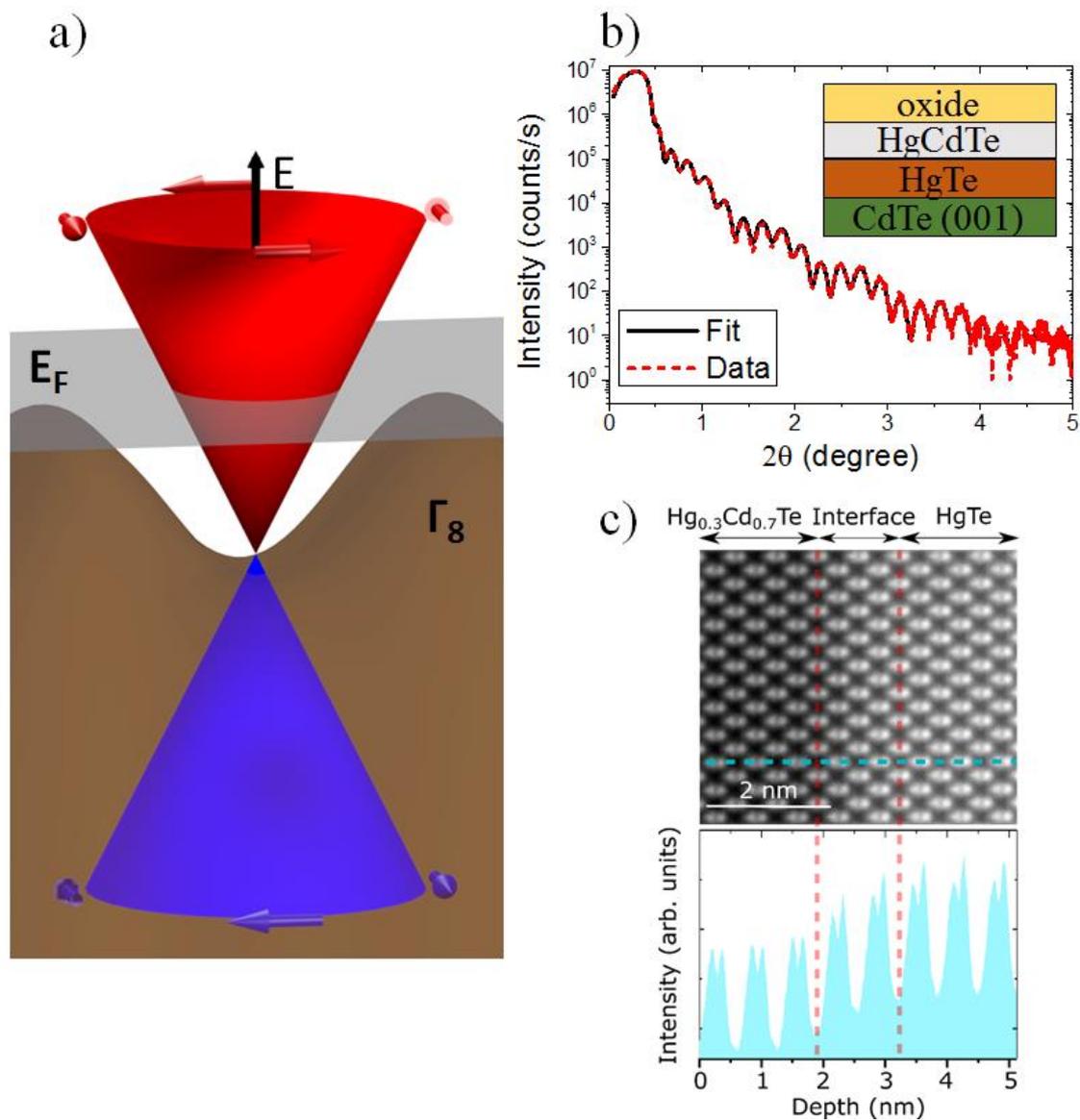

Figure 1: a) Schematic representation of the band structure of strained HgTe, with the Dirac dispersion cone of the surface states, and the bulk $\Gamma_8$ band. The arrows represent the helical spin configuration. b) X-Ray Reflectivity spectrum of a HgTe (18.5nm)/HgCdTe (5.5nm) sample. The structure used for the fit is represented in the inset. The red dashed curve represents the experimental data, the black curve is the fit. c) Scanning tunneling electron microscopy HAADF image of HgCdTe/HgTe/HgCdTe structure and corresponding chemical profile.

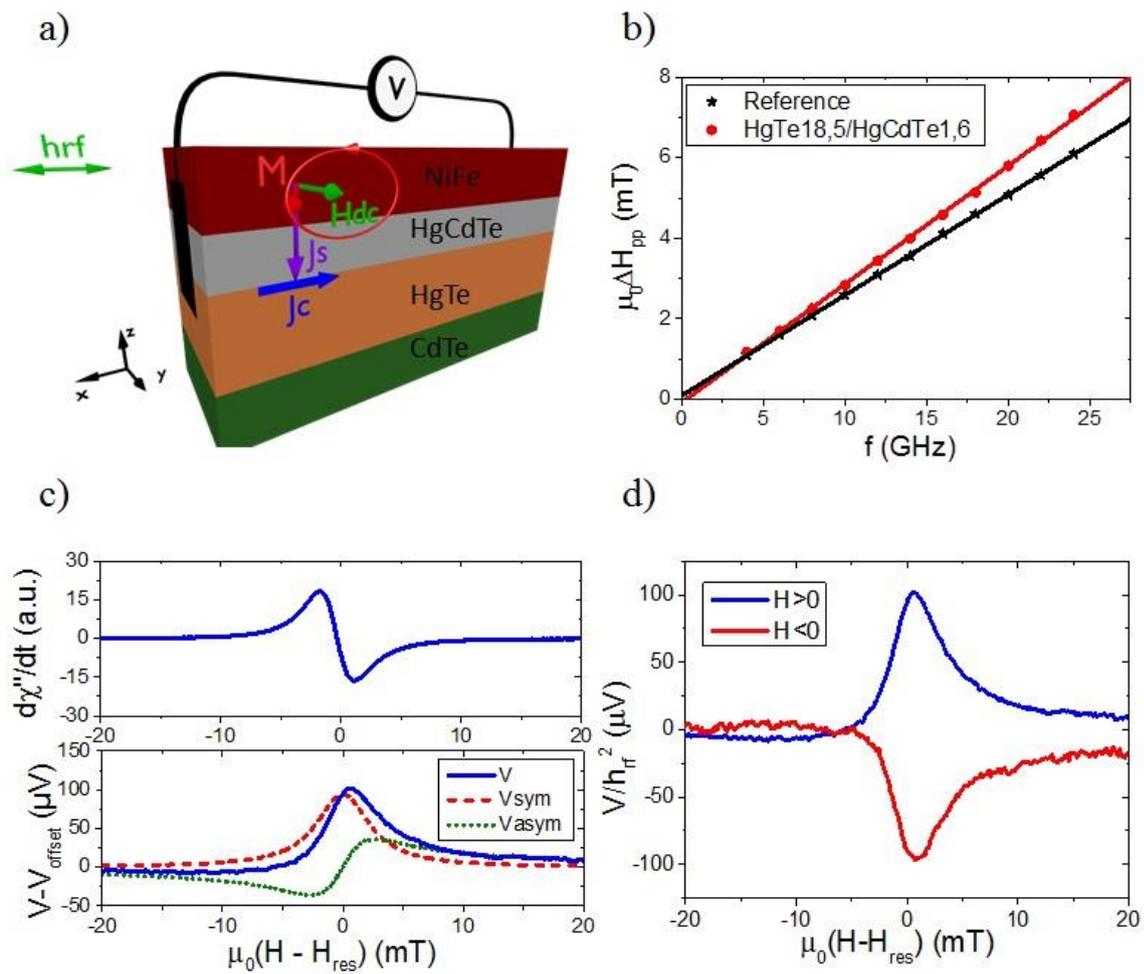

Figure 2: a) Principle of the spin pumping ferromagnetic resonance (FMR) measurement. b) Broadband frequency dependence of the peak-to-peak FMR linewidth of the reference NiFe/Si sample, and of a NiFe/HgCdTe$_{1.6}$/HgTe$_{18.5}$ sample. The damping coefficient of NiFe is higher when deposited on HgTe ($\alpha_{ref}$ =6.33x10$^{-3}$±3x10$^{-5}$ for the reference sample, compared to $\alpha$=7.50x10$^{-3}$±7x10$^{-5}$). For a *rf* field of 0.1 mT, this leads to a pure spin current $J_s^{3D}$=7.6±0.2MA/m². c) FMR and DC voltages, measured by spin pumping FMR on the same sample. The symmetric (red) and antisymmetric (green) contributions have been extracted from the measured signal (in blue). d) Spin-pumping signals obtained for a positive and a negative DC field, on the same sample.

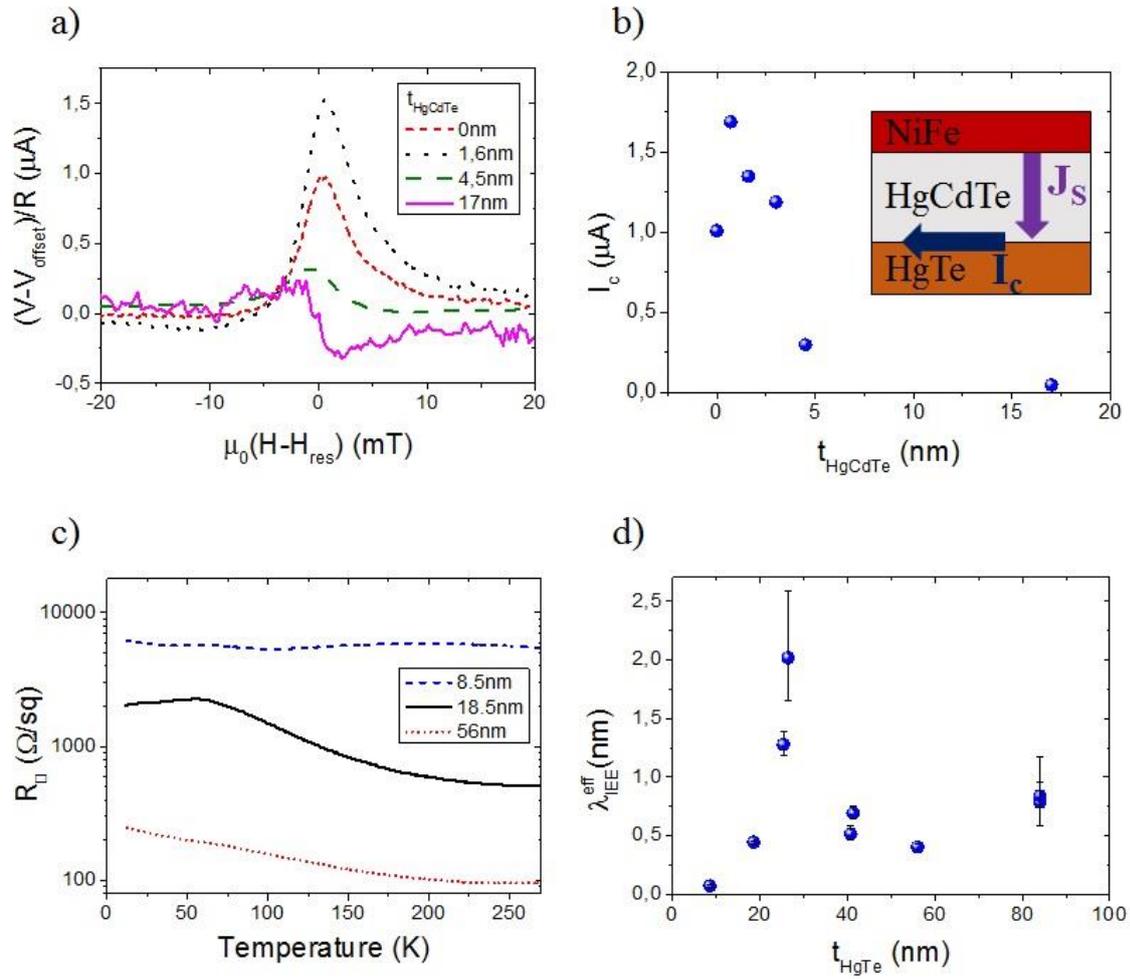

Figure 3: a) Spin pumping signals obtained for different thicknesses of HgCdTe barriers, normalized by the sample resistance. b) HgCdTe thickness dependence of $I_C$. The measurement were all performed on HgTe layers of the same thickness (18.5nm), while varying the HgCdTe layer thickness using Ar etching. The thicknesses were measured by XRR. Inset: scheme of the conversion. c) Sheet resistance R□ as a function of the temperature, for three samples of different HgTe thicknesses (8.5nm, 18.5nm and 56nm). d) HgTe thickness dependence of the inverse Edelstein length $\lambda_{IEE}$. The HgCdTe barrier thickness is the same for all the samples ($t_{HgCdTe}$~1.6 nm). The large error bars for the 26 nm and 84 nm thick samples are due to a relatively large uncertainty on the extra damping.